\documentclass{aastex62}

\submitjournal{ApJ}

\shorttitle{Intermittency out of the  Ecliptic Plane}
\shortauthors{Wawrzaszek et al.}
\begin{document}

\title{Multifractal Analysis of Heliospheric Magnetic Field Fluctuations observed by Ulysses}

\correspondingauthor{Anna Wawrzaszek}
\email{anna.wawrzaszek@cbk.waw.pl}

\author[0000-0001-9946-3547]{Anna Wawrzaszek}
\affil{Space Research Centre, Polish Academy of Sciences \\
Bartycka 18 A\\
00-716 Warszawa, Poland}

\author{Marius Echim}
\affiliation{Royal Belgian Institute for Space Aeronomy, \\
Brussels, Belgium}
\affiliation{Institute of Space Science,\\
M\u{a}gurele, Romania}
\email{marius.echim@oma.be}

\author{Roberto Bruno}
\affiliation{Institute for Space Astrophysics and Planetology,\\
Roma, Italy}
\email{roberto.bruno@inaf.it}

\begin{abstract}

We present the multifractal study of the intermittency of the magnetic field turbulence in the fast and slow solar wind beyond the ecliptic plane during two solar minima (1997-1998, 2007-2008) and solar maximum (1999-2001). More precisely, we consider 126 time intervals of \textit{Ulysses} magnetic field measurements, obtain the multifractal spectra, and examine the degree of multifractality as the measure of intermittency in the MHD range of scales, for a wide range of heliocentric distances and heliolatitudes. The results show a slow decrease of intermittency with the radial distance, which is more significant for the fast than for the slow solar wind. Analysis of Alfv\'{e}nic and compressive fluctuations confirms the decrease of intermittency with distance and latitude. This radial dependence of multifractality/intermittency may be explained by a slower evolution of turbulence beyond the ecliptic plane and by the reduced efficiency of intermittency drivers with the distance from the Sun. Additionally, our analysis shows that the greatest differences between magnetic field components are revealed close to the Sun, where intermittency is the strongest. Moreover, we observe that the slow solar wind from the maximum of the solar cycle 23 exhibits in general, a lower level of multifractality (intermittency) than fast solar wind, which can be related to the idea of a new type of Alfv\'{e}nic slow solar wind.

\end{abstract}

%magnetohydrodynamics (MHD) – methods: data analysis – solar wind – turbulence
%interplanetary medium  magnetohydrodynamics (MHD)  methods: data analysis  solar wind 
%Turbulence,intermittency, solar wind

\section{Introduction} 
\label{sec:intro}

Turbulence in the solar wind was analyzed extensively based on measurements acquired from various space missions (see~\citep{BruCar13} for review). A large number of studies focused on the existence of the intermittency and its role for the inertial range turbulence~\citep[e.g.,][]{Bur91a,MarLiu93,Tuet96,Soret99,Bruet03}, as well as for processes in the kinetic range~\citep{Peret12,Wanet12,Soret17}.
Intermittency was analyzed from the anomalous scaling of structure functions~\citep[e.g.,][]{Bur91a,Bur95, MarLiu93, HorBal97}, from the non-Gaussian behavior of probability distribution functions (PDFs) of the solar wind fluctuation~\citep{MarTu94}, by using Castaing distributions~\citep{Caset90,Soret99,Soret01,PagBal03} or the fourth moment (kurtosis) as a first indicator of intermittency~\citep[e.g.,][]{Bruet03,Yoret09}. Some authors quantified the intermittency from the local intermittency measure based on wavelet analysis~\citep{Far92,VelMan99,Bruet01} or from the multifractal description~\citep[e.g.,][]{MarTu96,BurNes10,WawMac10,Wawet15}. In general, the analysis showed that intermittency is a real characteristic of turbulence in the slow and fast solar wind, during solar minimum and solar maximum, in the inner and outer heliosphere~\citep[see, e.g.,][]{Bur91a,Bur91b, MarLiu93}, further in the heliosheath~\citep{BurNes10,Macet11}, and even near the heliopause~\citep{Macet14}. 

In particular,~\cite{MarLiu93} showed that small scales are more intermittent than large scales, the fast solar wind is generally less intermittent than slow wind. Moreover, the radial component of the fluctuations in the fast solar wind is more intermittent than the transverse components. A comparison between fast and slow streams at two heliocentric distances, namely at $0.3$ au and $1$ au, allowed authors to suggest also the radial evolution of intermittency in the ecliptic.

\cite{Tuet96} using \textit{Helios} data (3 cases from slow and 2 from fast solar wind) studied the intermittent nature of magnetic field fluctuations in the range of scales $81-2500$ s. They examined the structure function scaling and showed that while the intermittency in the fast solar wind is relatively stable, the intermittency parameter for the magnetic components in the slow wind varies significantly, from relatively low values to very high ones, but without the clear radial evolution trend.

\cite{Soret99} used the Castaing distribution to analyze the fluctuation observed by \textit{Helios} 2 at time scales from $81$ s to $1$ day and showed that  the magnetic field strength is more intermittent than bulk speed in the fast and slow streams. Moreover, they indicated that the intermittency of the magnetic field intensity was roughly the same for both slow and fast wind.

~\cite{Bruet03} analyzed the flatness of magnetic field and velocity PDF at $0.3$, $0.7$ and $0.9$ au and concluded that intermittency in the fast solar wind increases with the increase of the distance from the Sun, while slow wind intermittency does not show radial dependence. They also considered the intermittency of the components rotated into the mean field reference system and showed that the most intermittent magnetic component is the one along the mean field, while the perpendicular ones show a similar level of intermittency within the associated uncertainties. Moreover, when the radial distance increases, the component along the mean field becomes more and more intermittent compared to the transverse ones.

These results were obtained from the ecliptic plane data, where velocity shear, parametric decay, and interaction of Alfv\'{e}nic modes with convected structures plays an important role~\citep{BruCar13}. Beyond the ecliptic plane, the state of turbulence differs from that observed in the equatorial region~\citep{Ruzet95,Bavet00,Bavet01}. Various studies showed that the evolution of turbulence in the polar wind is less rapid~\citep[e.g.,][]{Horet96,BruTre14}. The state of the turbulence influences the intermittency level, as shown for interplanetary magnetic field fluctuations~\citep[e.g.,][]{Ruzet95,HorBal01,PagBal02,PagBal03,Yoret09}.

In particular,~\cite{PagBal01} analyzed structure function scaling and used several intermittency models to look at small-scale magnetic field fluctuations in the solar wind measured by \textit{Ulysses} during solar minimum (1994-1995), at radial distances $1.3$ and $2.4$ au (heliolatitudes varied from $80^{\circ}$S to $80^{\circ}$N). Concentrating on the range of scales between $10$ s and $300$ s they indicated the high level of intermittency in the fast solar wind and showed how much more variable and difficult to interpret the state of the slow solar wind is. Transverse components showed a very similar level of intermittency, while the radial component values were slightly lower.

In another study,~\cite{PagBal02} confirmed that magnetic field components present a high level of intermittency at minimum (1994-1996) and at maximum (2000-2001) phases of the solar cycle, and showed that the slow solar wind has a lower level of intermittency compared to the fast flow.
However, these authors used only the wind speed as the criterion to discriminate between pure coronal fast and typical equatorial slow wind.\\
~\cite{PagBal03} fitted the PDFs of magnetic field fluctuations with the Castaing distribution for time scales between $40$ and $200$ s. They analyzed 28 cases of the pure polar coronal fast wind at solar minimum, between 1994 and 1996, and showed that intermittency of the magnetic field fluctuations increased with increasing the radial distance in the range $1.4-4.1$ au. The same authors showed that, in the RTN reference system, the transverse magnetic field components are significantly more non-Gaussian than the radial component. 

\cite{Yoret09} performed a detailed selection of the four states of the solar wind (pure fast, fast streams, pure slow, and slow streams) measured by \textit{Ulysses} during 1992-1997. In summary, the authors considered $21$ time intervals of length varied from $2$ to $19$ days. Their analysis of flatness performed by using $20$ s averages showed that only pure fast wind presents evolution with the heliocentric distance in the considered range between $1.5$ and $3.0$ au and at latitudes between $50^{\circ}$ and $80^{\circ}$. However, it is worth stressing that this evolution has been identified on the basis of four 4 cases only~\citep[][Table 1]{Yoret09}.~\citeauthor{Yoret09} concluded that pure slow wind measured at $5.1-5.4$ au and at latitudes narrowed to $<20^{\circ}$ presents the most intermittent state.

~\cite{Wawet15} performed multifractal analysis of fluctuations of the interplanetary magnetic field strength for two solar minima (1997-1998, 2007-2008) and one maximum (1999-2001). Their data included 98 time intervals of the slow and fast solar wind in the range of radial distances $1.4-5.4$ au and a heliolatitude range between $-80^{\circ}$ and $70^{\circ}$. It was found, in general, that the level of intermittency decreases with distance and latitude. It is worth pointing out, that during the identification of the multifractal scaling region, scales less than $16$ s have also been included, which could influence on the obtained results.

Comparative studies of solar wind intermittent turbulence beyond the ecliptic plane, especially at the solar minimum and the solar maximum, both for fast and slow solar winds, are limited and we aim to fill this gap in this work. We are convinced that the further extensive studies of the intermittency phenomenon in the solar wind by using the same methodology and time scales are still needed, especially in the context of the better understanding of the relation of the fluctuations' power within the inertial range with their nature of scaling below this range~\citep{Bruet14}. Therefore, we extend our previous studies by using a larger collection of magnetic field data sets (in summary $126$ time intervals), which provides a good statistical significance and allows us to examine the evolution of the degree of multifractality as a measure of intermittency over a wide range of radial distances and heliographic latitudes explored by \textit{Ulysses} spacecraft. To avoid the problem of mixing of different states of the solar wind, a strict procedure of data selection is applied. This is for the first time that the multifractal spectrum has been determined for three components of the interplanetary magnetic field in MHD range (on scales higher than $16$ s) at solar minima (1997-1998, 2007-2008) and maximum (1999-2001). Moreover, by using multifractal description we also investigate the intermittency of the parallel and perpendicular component in the local Mean Field reference system, as done by~\cite{Bruet03} for \textit{Helios} 2 data in the ecliptic plane. \\
The procedure for data selection is described in Section~\ref{sec:web:ud}. In Section~\ref{sec:web:ma} we briefly discuss the multifractal methodology used to quantify the intermittent state of considered data. Section~\ref{sec:web:res} presents the results of the multifractal analysis, showing in particular the latitudinal and radial dependences of level of multifractality as a measure of intermittency. The results for all magnetic field components, independently on the used reference system, reveal a slow decrease of the level of intermittency  with increasing distance from the Sun. Moreover, the slow solar wind measured during maximum of the cycle 23 presents a level of intermittency smaller than fast solar wind during the solar minimum of the same cycle. A summary is provided in Section~\ref{sec:web:con}.

\section{Data Selection Process}
\label{sec:web:ud}

In this study, we focus on the analysis of the magnetic field fluctuations measured by the \textit{Ulysses} spacecraft in the slow and fast solar wind. We analyze data from two solar minima (1997-1998, 2007-2008) and one solar maximum (1999-2001). We discriminate between the fast and slow solar wind state based on data selection criteria constructed with five solar wind parameters: (1) the radial velocity, $V_R$, (2) the proton density, $n_p$, (3) the proton temperature $T_p$, parameters provided by the Solar Wind Observations Over the Poles of the Sun instrument \citep{Bamet92}, (4) the oxygen ion ratio ($\mathrm{O}^{+7}/\mathrm{O}^{+6}$) from the Solar Wind Ion Composition Spectrometer~\citep{Gloet92}, and (5) the magnetic compressibility factor $C_B$, defined as  $C_B=(\langle|B|^2\rangle-\langle|B|\rangle^2)/((\langle B_{R}^{2}\rangle-\langle B_{R}\rangle^{2})+(\langle B_{T}^{2}\rangle-\langle B_{T}\rangle^{2})+(\langle B_{N}^{2}\rangle-\langle B_{N}\rangle^{2}))$~\citep{BruBav91}. $B_R$, $B_T$, and $B_N$ are the magnetic field components measured by the VHM-FGM magnetometer~\citep{Balet92}, while $|B|$ is the magnetic field strength. 
The five parameters listed above describe basic differences between slow and fast wind: the slow solar wind contrary to the fast wind is characterized by low speed, stronger magnetic compressibility, higher density, and lower proton temperature~\citep{Yoret09,BruCar13,Lepet13}.
We selected the oxygen charge states ($\mathrm{O}^{+7}/\mathrm{O}^{+6}$) as a diagnostic for the origin 
different solar wind type~\citep{Zuret02,Yoret09,Lanet12}.
~\cite{Lanet12} showed that the carbons ratio $\mathrm{C}^{+6}/\mathrm{C}^{+4}$ can be considered as an even better indicator of solar wind type and region of origin. We also tested this parameter and found that it gives similar results as $\mathrm{O}^{+7}/\mathrm{O}^{+6}$. 
For each of the $5$ parameters we defined a threshold value for the transition from slow to the fast wind. The threshold values, listed in Table~\ref{t:web:1}, have been determined empirically by visual inspection of 6 hr averages of each solar wind parameter.

\begin{table}[!h]
\caption{The threshold Values Applied to the Five Parameters Used to Select Slow and Fast Solar Wind Streams in Each Year of Three Periods of Solar Activity.}
\begin{center}
\begin{tabular}{|c|c|c|c|c|c|c|c|}
\hline
\textbf{Threshold} & \multicolumn{2}{c|}{\textbf{Solar Minimum}} & \multicolumn{3}{c|}{\textbf{Solar Maximum}} & \multicolumn{2}{c|}{\textbf{Solar Minimum}}\\
\cline{2-8}
                                     &  $1997$ & $1998$ & $1999$ & $2000$& $2001$   & $2007$ & $2008$\\
\cline{2-8}
                                     &  $4.7-5.4$ au & $5.2-5.4$ au & $4.2-5.2$ au & $2.0-4.2$ au & $1.3-2.6$ au   & $1.4-2.6$ au & $2.0-4.1$ au\\
\hline
$V_{\mathrm{thr}}$ (km s$^{-1}$)                   & $500$  & $450$  & $450$  & $450$  & $500$    &  $500$  & $500$ \\
\hline
$(\mathrm{O}^{+7}/\mathrm{O}^{+6})_{\mathrm{thr}}$  &  $0.1$ & $0.1$ &$0.1$ & $0.1$  &$0.1$ & $0.05$& $0.05$ \\
\hline
${C_B}_{\mathrm{thr}}$                              &   $0.1$ & $0.1$ &$0.1$ & $0.1$ & $0.1$  & $0.1$& $0.1$ \\
\hline
$T_{\mathrm{thr}}$  $(K)$                                &  $5 *10^4$ (d\footnote{d denotes the day of the year} $<160$)& $4*10^4$ &$5 *10^4$ & $5 *10^4$ & $1 *10^5$  & $8 *10^4$& $8 *10^4$ \\
                                                    &  $4 *10^4$ (d$\geq 160$)& & &  &   & &  \\
\hline
$n_{\mathrm{thr}}$  (cm$^{-3}$)                                 &   $0.2$ & $0.2$ &$0.2$ & $0.4$ (d$>200$)& $1.5$ & $1.5$ & $0.7$ (d$<200$)\\
                                                    &    &  & &   $1.2$ (d$\geq 200$) & & & $0.3$ (d$\geq 200$)\\
                                                    \hline
$n_{\mathrm{thr}}$ (cm$^{-3}$)                                 &   $5.2$ & $5.7$ &$4.5$ & $4.9$ (d$>200$)& $5.7$ & $6.0$ & $5.1$ (d$<200$)\\
(at 1 au)                                            &    &  & &   $7.5$ (d$\geq 200$) & & & $4.2$ (d$\geq 200$)\\
\hline
\end{tabular}
\end{center}
\label{t:web:1}
\end{table}
The variation of the threshold values in Table~\ref{t:web:1} reflects the motion of \textit{Ulysses} at different radial distances and heliolatitudes~\citep{Yoret09,BruCar13,Lepet13}. Moreover, quantities in Table~\ref{t:web:1} vary also with the solar cycle~\citep[e.g.,][]{Kaset12}. The threshold values chosen for $\mathrm{O}^{+7}/\mathrm{O}^{+6}$ at minimum (1997-1998) and maximum (1999-2001) are in full agreement with the previous study~\citep{Zuret02}. However, for the solar minimum (2007-2008) we have identified thresholds $(\mathrm{O}^{+7}/\mathrm{O}^{+6})_{\mathrm{thr}}$ two times smaller, which independently confirms the unusual character of the solar wind during this time interval and possibly significant changes occurring globally on the Sun (for more details, see, e.g. \citep{McCet08,Lepet13}).
In the final step of data selection we excluded time intervals corresponding to interplanetary transients like shocks and CMEs, identified by Gosling and Forysth (see http://www.sp.ph.ic.ac.uk/Ulysses/shocklist.txt) and Gosling and Ebert (see http://swoops.lanl.gov/cme\_list.html). \\
Based on the procedure reported above we were able to select 126 time intervals, of which $88$ correspond to fast solar wind ($38$ cases during solar maximum and $50$ during minimum) and $38$ time intervals to slow solar wind ($28$ during solar maximum and $10$ during minimum). Each selected data set contained a few days long time series of the radial ($R$), transversal ($T$), and normal ($N$) component of the magnetic field measured in RTN system with resolution $1-2$ vectors S$^{-1}$. In the RTN system, the $R$ axis is along the radial direction, positive from the Sun to the s/c, the $T$ component is perpendicular to the plane formed by the rotation axis of the Sun $\Omega$ and the radial direction ($T=\Omega\times R$), and the $N$ component resulting from the vector product $N=R\times T$~\citep{BruCar13}.
Moreover, in order to avoid the problem of cross-talking between the components and to provide a complete picture of the radial dependence, all data have been transformed to the mean field (hereafter MF) coordinate system. The description of the MF system can be found in Appendix D.2 of the review by~\cite{BruCar13}, see also~\cite{Bruet03}. In this reference system one of the components, $B_{\parallel}$, is parallel to the mean field $\textbf{B}_0$ and outwardly oriented, the $B_{\perp 2}$ is perpendicular to the plane containing $\textbf{B}_0$ and the $\textbf{R}$ axis in the RTN system, while $B_{\perp 1}$ is the result of vector product $\textbf{B}_{\perp 1}=\textbf{B}_{\perp 2}\times \textbf{B}_0$ (here $\textbf{B}_{\perp 1}$ and $\textbf{B}_{\perp 2}$ indicate a unit vector). 
  
\section{Multifractal Analysis}
\label{sec:web:ma}

The state of intermittency in the solar wind can be described by using several methods. In this work we apply a multifractal formalism~\citep[e.g.,][]{Halet86,Fal90}, in the frame of which both large and small concentrations of measure are taken into account, which
seems to be the most appropriate to analyze the complex state of the small-scale intermittent turbulence. Multifractals can be described by functions and group of parameters~\citep[e.g.,][]{GraPro83,Halet86,Fal90,Ott93}. One of the basic characteristics of multifractal scaling is the multifractal spectrum $f(\alpha)$, sketched in Figure~1 of the paper by \cite{WawMac10}. As a function of a singularity strength $\alpha$, $f(\alpha)$ portrays the variability in the scaling properties of the measures. The shape of $f(\alpha)$, in particular its width is used to quantitatively estimate the degree of multifractality of considered data~\citep[e.g.,][]{Ott93, MacWaw09,WawMac10,Macet14}.

The multifractal spectrum can be determined with different methods~\citep{GraPro83,ChhJen89,Kanet02}. In this study we apply the partition function (PF) technique~\citep{Halet86}, successfully tested in many situations for space plasma turbulence~\citep[e.g.,][]{MacSzc08,BurNes10,Lamet10,Wawet15}. 
In the first step of the  PF analysis, the multifractal measure is constructed by using the first moment of increments of magnetic fluctuations,
\begin{equation}
\epsilon(x_i,l)=\mid B(x_i+l)-B(x_i )\mid  
\end{equation}
where $i = 1, \ldots, N = 2^{n}$, with $n\geq 17$ denotes data set length as a power of 2, while $B$ denotes one of the magnetic field components ($B_R$, $B_T$, $B_N$ in RTN or $B_{\parallel}$, $B_{\perp 1}$, $B_{\perp 2}$ in MF reference system), separated from a position $x_i$ by a distance $l$. It is worth pointing out that different measures can be constructed to emphasize various effects and to describe physical processes in data~\citep{Man89,Bur95,WawMac10,Soret17}. Next, we divide the signal into segments of size $l$ and then each $i$th segment is associated with a probability measure defined by
\begin{equation}
p(x_i,l)\equiv \frac{\epsilon(x_i,l)}{\sum_{i=1}^{N}\epsilon(x_i,l)}=p_i(l).
\label{e:web:measure}
\end{equation}
This quantity can be interpreted as a probability that the portion of fluctuation is transferred to an eddy of size $l$. As usual, at a given position $x=v_{\mathrm{sw}} t$, where $v_{\mathrm{sw}}$ is the average solar wind speed, the temporal scales measured in units of sampling time $\Delta t$ can be interpreted as the spatial scales $l=v_{\mathrm{sw}} \Delta t$ if the Taylor frozen-in-flow assumption~\citep{Tay38} is satisfied~\citep{Howet14,Per17}.
Next, the partition function $\chi (q,l)$ for various values of scales $l$ and moments $q$ is defined as follows:
\begin{equation}
\chi (q,l)=\sum_{i=1}^{N(l)}(p_{i} (l))^q.
\label{e:web:pf}
\end{equation}
This definition allows us to take into account both large ($q>0$) and small ($q<0$) concentrations of measure, in contrast to a standard statistical description, where only positive moments are considered. Moreover, analyzing the PF (see Equation \ref{e:web:pf}), we can control and describe how moments $q$ of measure (Eq. ~\ref{e:web:measure}) scale with $l$. 
In particular, for a multifractal measure, this PF scales with the box size $l \rightarrow  0$ as: 
\begin{equation}
\chi (q,l)\propto l^{\tau(q)},
\label{e:web:spf}
\end{equation}
where $\tau(q)$ is the scaling exponent determined in the further step as the slopes of the logarithm of $\chi (q,l)$ ( Eq.~\ref{e:web:spf}) versus $\log l$.  The multifractal scaling range, where the linear regressions can be applied and $\tau(q)$ determined, is not known a priori, and it must be chosen for each case separately. We emphasize here that to locate the optimal scaling range an automatic selection procedure proposed by~\cite{SauMul99} has been used. Previous studies showed that even very small scales from the dissipation region can present multifractal scaling~\citep[e.g.,][]{Soret17}. The high frequency limit/break of the inertial range changes with the radial distance~\citep{BruTre14}: is around $0.2/0.1/0.07$ Hz at $1.4/3.2/5.3$ au. Therefore, multifractal analysis has been limited to scales not smaller than $16$ s.\\
In the next step of analysis, the function $f(\alpha)$ is determined through a Legendre transformation $\alpha(q)=\frac{d}{dq}[\tau(q)]=\tau'(q)$, $f(\alpha(q))= q\alpha(q)-\tau(q)$~\citep{Halet86,ChhJen89}. The multifractal spectrum ($\alpha$, $f(\alpha)$) is compared with the results of an independent methodology proposed by~\cite{ChhJen89}, in which $\alpha$ and $f(\alpha)$ can be obtained directly from the slopes of generalized measures~\citep[e.g.][]{MacWaw09,WawMac10,Wawet15}.
Finally, the width of the singularity spectrum determined by the difference
\begin {equation}
\Delta\equiv \alpha_{\mathrm{max}} -\alpha_{\mathrm{min}},
\label{e:web:dm}
\end{equation}
where $f(\alpha_{\mathrm{min}})=f(\alpha_{\mathrm{max}})=0$, quantifies the degree of multifractality of considered data~\citep[e.g.,][]{MacSzc08}. In general, the parameter $\Delta$ describes the level of inhomogeneity and is naturally related to the deviation from a strict self-similarity. That is why $\Delta$ is also a measure of intermittency, which is in contrast to self-similarity~\cite[][chapter 8]{Fri95}. 
In practice, owing to the limited data set, only values close to the maximum of the function $f(\alpha)$ can be calculated. Therefore, in the final step of analysis, the values $\alpha_{\mathrm{max}}$ and $\alpha_{\mathrm{min}}$ are fitted with a p-model for symmetric distributions~\citep{MenSre87} and a two-scale binomial multiplicative cascade model for asymmetric spectra~\citep{MacSzc08}. This approach was successfully used in many studies of space plasma turbulence~\citep{Buret93,MacSzc08,SzcMac08,MacWaw09,BurNes10,Lamet10}. In our previous work~\citep[][Figure 1]{Wawet15} we show an example of such a fit. It is worth underlining that the compatibility of the spectrum $f(\alpha)$ with the phenomenological intermittent multiplicative models confirms that multifractality at small scales is the result of intermittent nonhomogeneous cascades. Moreover, the asymmetry of the multifractal spectra can be used as an additional parameter for describing the intermittent nature of the fluctuations in the solar wind~\citep[e.g.,][]{Macet11,Macet14}. 

\section{Results}
\label{sec:web:res}

The procedure presented above was applied to compute the multifractal spectra $f(\alpha)$ for $126$ time intervals selected during two solar minima (1997-1998, 2007-2008) and one maximum (1999-2001). Next, we performed a quantitative analysis on how the widths of these spectra $f(\alpha)$ change; we determined the degree of multifractality $\Delta$ (see Equation~\ref{e:web:dm}). Figures~\ref{f:web:radRTN}, ~\ref{f:web:latRTN}, and ~\ref{f:web:map_RTN} present the values of the parameter $\Delta$ as a measure of intermittency computed for  fast (left column) and slow (right column) solar wind and the three magnetic field components $B_R$, $B_T$ and $B_N$. We also indicate the accuracy of the calculation of the parameter $\Delta$. For comparison, in Figures~\ref{f:web:radMFRS},~\ref{f:web:latMFRS}, and ~\ref{f:web:map_MFRS}, we present the results of multifractal analysis preformed for magnetic field components transformed to the MF reference system: $B_{\parallel}$ (parallel to the mean field vector $\textbf{B}_0$, panels (a) and (b)), and the two perpendicular components $B_{\perp 1}$ (panels (c) and (d)), $B_{\perp 2}$ (panels (e) and (f)). In the next subsections we discuss the radial and latitudinal dependences of multifractality as a measure of intermittency.

\subsection{Variation of Inertial Range Intermittency with Radial Distance}
\begin{figure}[!h]
\centering
\includegraphics[scale=0.9]{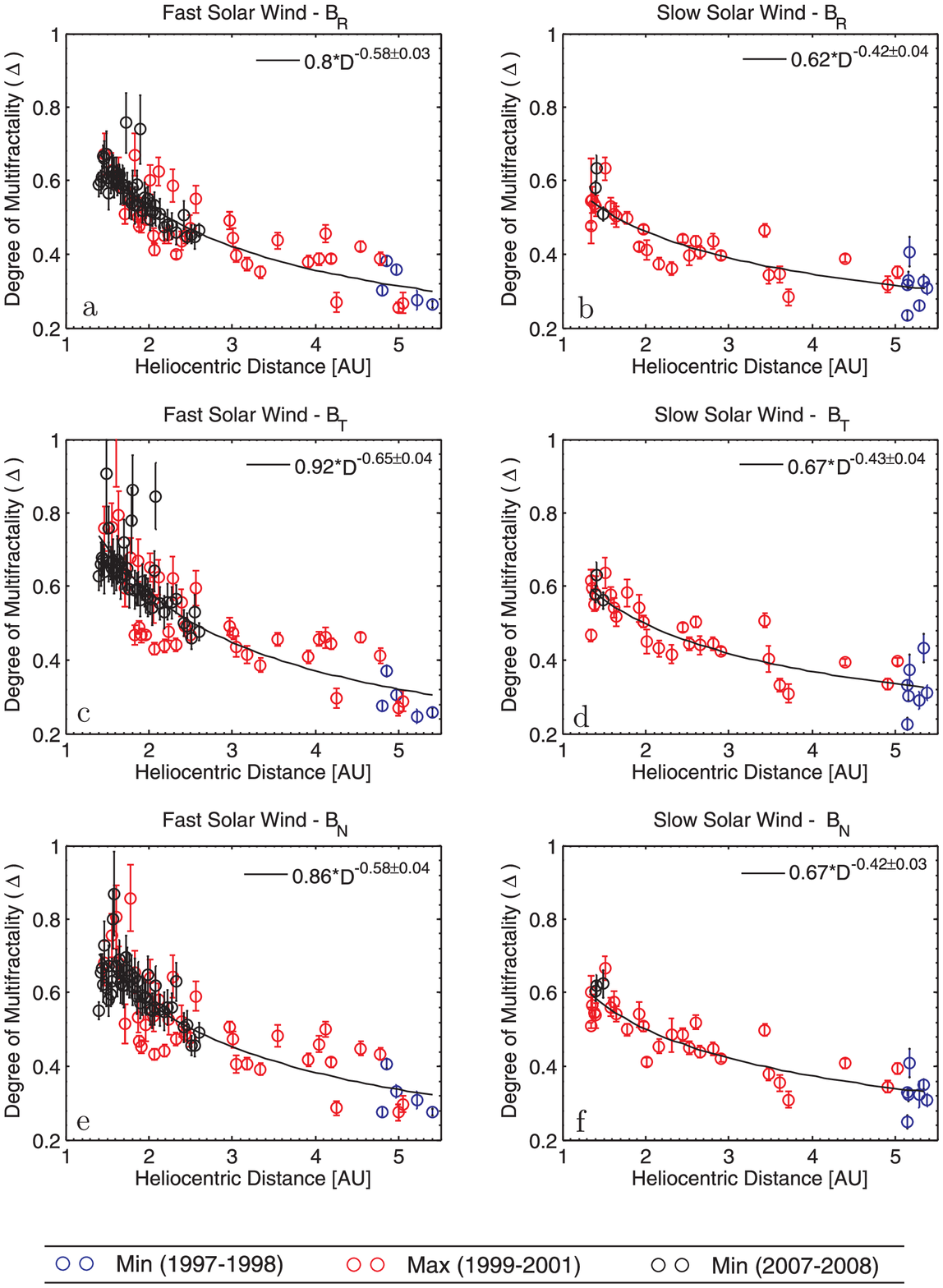}
 \caption{Degree of multifractality $\Delta$ as a measure of intermittency in radial $B_R$ (panels (a) and (b)), transversal $B_T$ (panels (c) and (d)) and normal $B_N$ (panels (e) and (f)) magnetic field components as a function of the distance from the Sun. Data were collected during two solar minima (1997-1998, blue circles) and (2007-2008, black circles), and one solar maximum (1999-2001, red circles). Power-law fit $\Delta=aD^{b}$ to data is presented by a~continuous line.}
 \label{f:web:radRTN}
 \end{figure}

\begin{figure}[!h]
\centering
\includegraphics[scale=0.9]{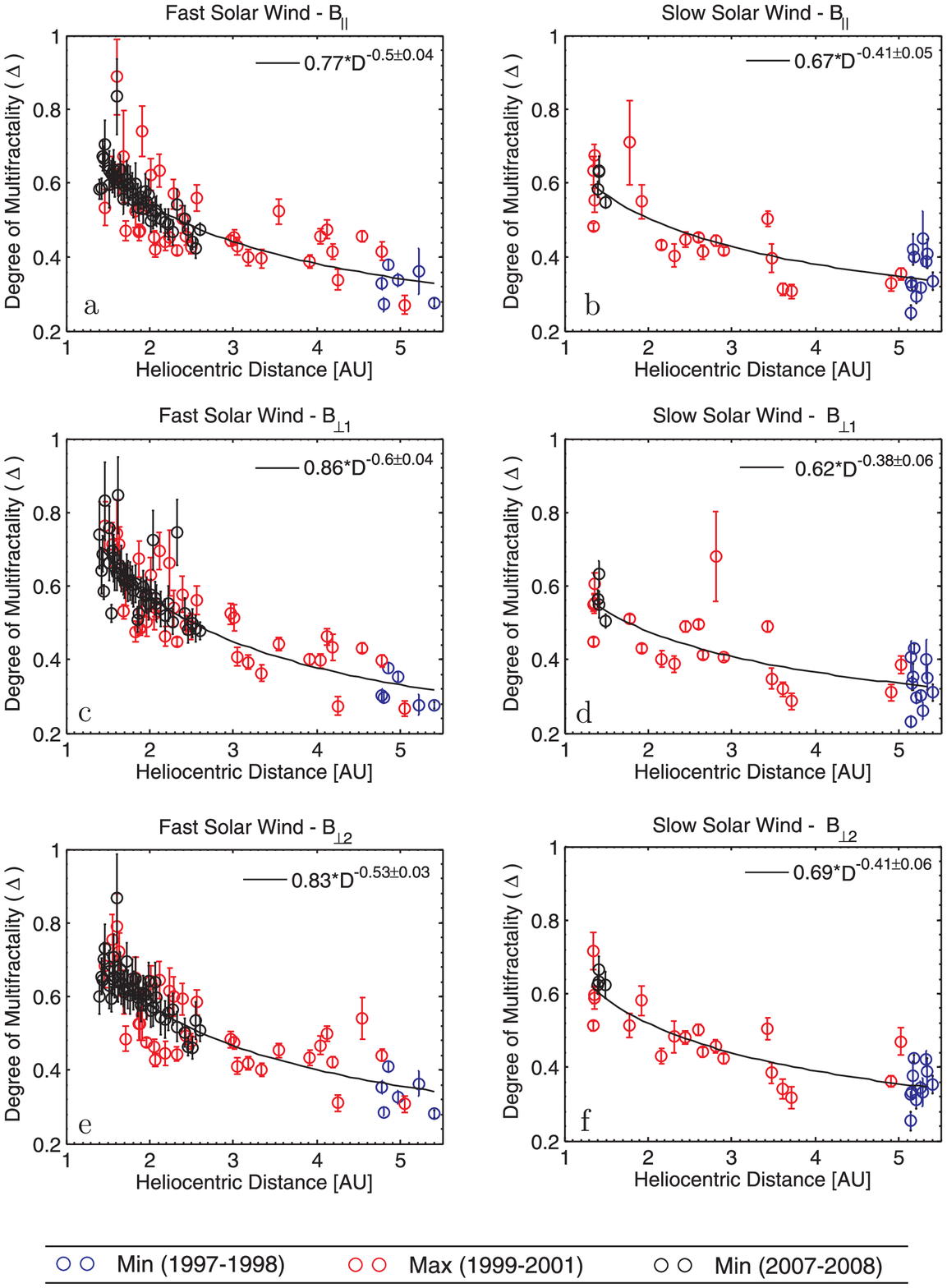}
 \caption{Degree of multifractality $\Delta$ as a measure of intermittency in magnetic field components measured during solar minima (1997-1998, 2007-2008) and solar maximum (1999-2001) and transformed to the MF reference system: $B_{\parallel}$ (parallel to the mean field vector $\textbf{B}_0$, panels (a) and (b)), and two components $B_{\perp 1}$ (panels (c) and (d)), $B_{\perp 2}$ (panels (e) and (f)), perpendicular to the field. Values of $\Delta$ presented as a function of the distance from the Sun together with the power-law fit $\Delta=aD^{b}$ to data marked by a~continuous line.}
 \label{f:web:radMFRS}
 \end{figure}

Figures~\ref{f:web:radRTN} and~\ref{f:web:radMFRS} show the variation of parameter $\Delta$ with the radial distance $D$, between $1.4$ and $5.4$ au from the Sun, in the fast and slow solar wind. The radial (Figure~\ref{f:web:radRTN}(a)), transversal (Figure~\ref{f:web:radRTN}(c)) and normal (Figure~\ref{f:web:radRTN}(e)) magnetic field components in the fast solar wind show a small decrease of intermittency with the distance from the Sun. The decreasing rate, determined by the least squares linear fit to the log-log values of $\Delta$ against distance $D$ ($\Delta=aD^{b}$), differ between components. More precisely, for $B_R$ and $B_N$ we found $b=0.58\pm 0.03$ while for $B_T$ the rate is slightly higher, $b=0.65\pm 0.03$. However, at heliocentric distances smaller than $2.5$ au we observe that the trend of variation departs slightly from this power-law regime.
The analysis of data transformed to MF reference system (Figure~\ref{f:web:radMFRS}) reveals similar behavior as that found for the RTN reference system. Again we observe the decrease of $\Delta$ with distance from the Sun, both for the parallel ($B_{\parallel}$) and perpendicular ($B_{\perp 1}$ and $B_{\perp 2}$) components. The smallest rate was found for $B_{\parallel}$, for which we have $b=0.5\pm 0.04$. For perpendicular components the parameter $b$ varies from $0.53\pm 0.03$ (for $B_{\perp 2}$) to $0.6\pm 0.04$ (for $B_{\perp 1}$). 

The results for the slow solar wind shown in Figures~\ref{f:web:radRTN} and~\ref{f:web:radMFRS} (panels (b), (d) and (f)) are more difficult for the interpretation. We analyzed a smaller number of samples, most of them (marked by red circles), obtained from the maximum of solar cycle 23. Additionally, in a few cases the lack of a clear multifractal scaling did not allow us to determine the parameter $\Delta$. Nevertheless, we can still observe a decrease of multifractality (intermittency) with radial distance but weaker than in the case of the fast solar wind. The parameter $\Delta$ for slow solar wind data in the RTN reference system decreases with distance $D$ with rates of about $0.42\pm 0.03$ (Figure~\ref{f:web:radRTN}(b), (d) and (f)). We obtained a similar behavior for components rotated into the MF reference system, as depicted in Figure~\ref{f:web:radMFRS}(b), (d) and (f), with $b$ in the range between $0.38\pm 0.06$ (for component $B_{\perp 1}$) and $0.41\pm 0.05$ (for $B_{\parallel}$ and $B_{\perp 1}$). The observed difference between components and reference systems seems to be rather negligible, particularly if we take into account the errors of the fit $\Delta=aD^{b}$, which are higher for the MF reference system.

Additionally, in order to compare with our previous analysis~\citep{Wawet15}, we performed a study, in which we included additional smaller scales, between $2$ and $16$ s. The rates obtained from power-law fit $\Delta=aD^{b}$ are summarized in Table~\ref{t:web:2} and compared to the ones corresponding to Figures~\ref{f:web:radRTN} and~\ref{f:web:radMFRS}.

 \begin{table}[!h]
\caption{The Variation Rate, $b$, of Multifractality (Intermittency) with the Radial Distance $D$ ($\Delta=aD^{b}$) for the Fast and Slow Solar Wind Streams in RTN and MF Reference Systems. The Table Compares Results from Multifractal Analysis at MHD Scales ($\geq 16$ s) and Kinetic Scales ($\geq 2$ s).}
\begin{center}
\begin{tabular}{|c|c|c|c|c|}
\hline
 & \multicolumn{2}{c|}{\textbf{Fast Solar Wind}}& \multicolumn{2}{c|}{\textbf{Slow Solar Wind}} \\
 \hline
Scale& $\geq 16$ s & $\geq 2$ s & $\geq 16$ s&  $\geq 2$ s  \\
\hline
$B_R$  &$0.50\pm 0.03$ &$0.39\pm 0.03$ &$0.42\pm 0.04$ &$0.20\pm 0.04$\\
$B_T$ & $0.65 \pm 0.04$&$0.39\pm 0.04$&$0.43\pm 0.04$ &$0.27\pm 0.04$\\
$B_N$ & $0.58 \pm 0.04$&$0.37\pm 0.03$&$0.42\pm 0.03$ &$0.32\pm 0.03$\\
\hline
$B_{\parallel}$&$0.50\pm 0.04$ &$0.15\pm 0.06$ &$0.41\pm 0.05$ &$0.10\pm 0.05$ \\
$B_{\perp 1}$&$0.60\pm 0.04$ &$0.24\pm 0.05$ &$0.38\pm 0.06$&$0.22\pm 0.06$ \\
$B_{\perp 2}$ &$0.53\pm 0.03$ &$0.29\pm 0.04$ &$0.41\pm 0.06$& $0.15\pm 0.07$\\
\hline
\end{tabular}
\end{center}
\label{t:web:2}
\end{table}

Table~\ref{t:web:2} shows smaller values of $b$ when scales smaller than $16$ s are included in the multifractal analysis. Similar rate values ($0.31\pm 0.03$ for fast solar wind and $0.26\pm 0.04$ for slow solar wind) have been found for multifractal analysis of magnetic field intensity~\citep[][Figure 2]{Wawet15}. This slower decrease of multifractality (intermittency) with distance can be somehow related to the higher level of intermittency at smaller scales, as confirmed by various studies~\citep[e.g.,][]{MarLiu93,Soret99,Bruet03,Alete08}. Another interesting finding is that the intermittency of the parallel component $B_{\parallel}$ shows a negligible variation with distance, both in the fast and slow solar wind. We suggest that it can be due to the fact that compressive phenomena, more important in $B_{\parallel}$~\citep{Bruet03}, start to be more active on scales smaller than $16$ s.~\cite{Alete08} showed that intermittency increases at kinetic scales possibly due to compressive effects.\\
On the basis of these results, we can tentatively conclude that the analysis of \textit{Ulysses} data, mainly from the solar cycle 23, reveals a decrease of intermittency with increasing distance from the Sun. However, the rate of variation, is always higher for fast than for slow solar wind and presents significant anisotropy between the three components of the magnetic field. The analysis shows that the smallest scales have a crucial role in determining the rate of radial variation of intermittency.

\subsection{Latitudinal Variation of Inertial Range Intermittency}
\label{sub:web:lat}
\begin{figure}[!htbp]
 \centering
\includegraphics[scale=0.9]{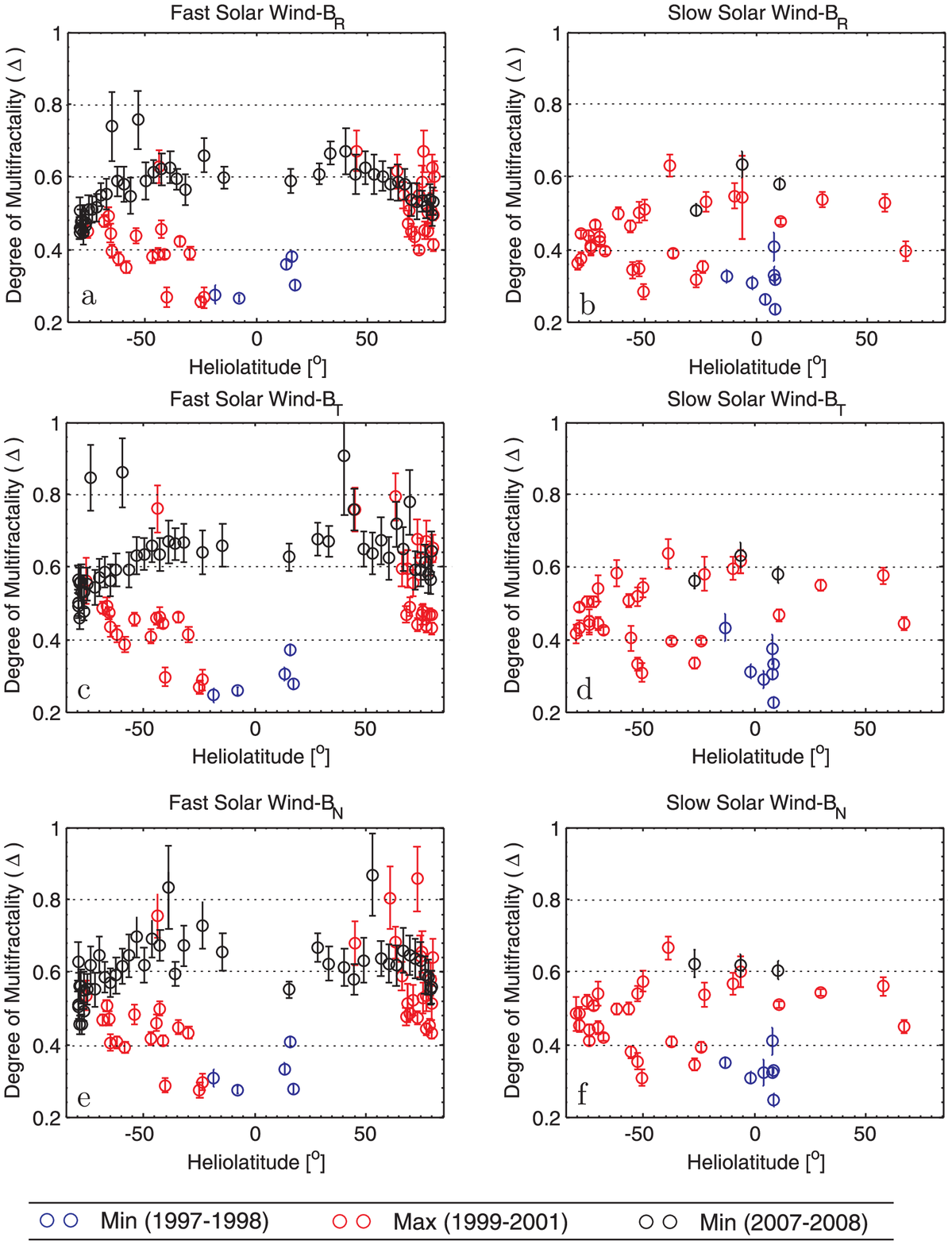}
 \caption{Parameter $\Delta$ quantifying multifractality (intermittency)
in the magnetic field components $B_R$ (a), (b), $B_T$ (c), (d), $B_N$ (e), (f) measured in RTN reference system as a function of the heliographic latitude during solar minima (1997-1998, 2007-2008) and solar maximum (1999-2001), respectively.}
 \label{f:web:latRTN}
 \end{figure}

 \begin{figure}[!htbp]
 \centering
\includegraphics[scale=0.9]{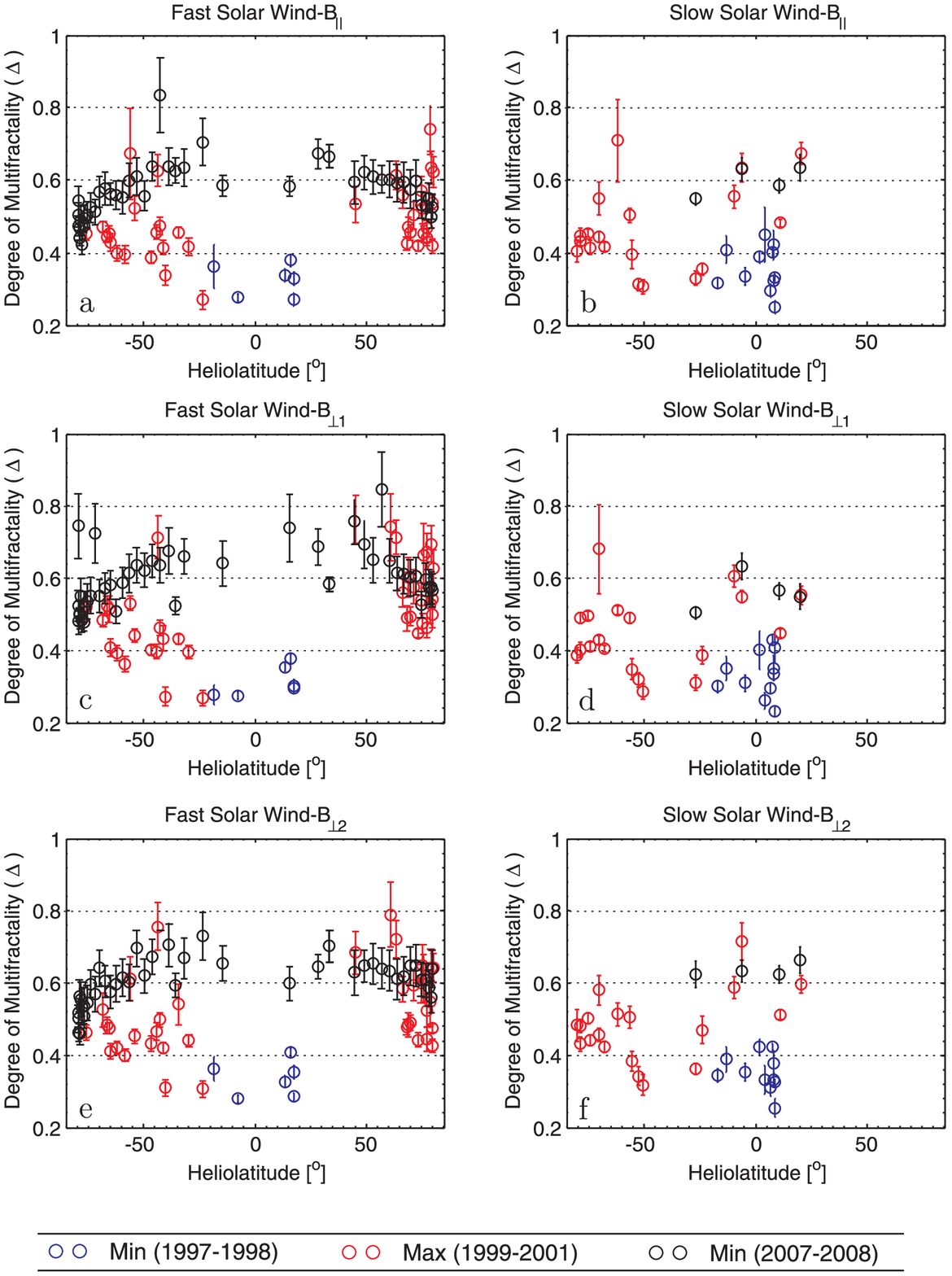}
 \caption{Latitudinal dependence of the parameter $\Delta$, quantifying multifractality (intermittency) in magnetic field components measured during solar minima (1997-1998, 2007-2008) and solar maximum (1999-2001). Data were transformed to the MF reference system: $B_{\parallel}$ (parallel to the mean field vector $\textbf{B}_0$, panels (a) and (b)), and two components $B_{\perp 1}$ (panels (c) and (d)), $B_{\perp 2}$ (panels (e) and (f)), perpendicular to the field, respectively.}
 \label{f:web:latMFRS}
 \end{figure}
 
The latitudinal dependence of the degree of multifractality $\Delta$ as a measure of intermittency for the fast and slow solar wind cases are depicted in Figures~\ref{f:web:latRTN} (multifractality of RTN components) and~\ref{f:web:latMFRS} (multifractality of MF components). For all cases of fast solar wind at solar minimum, we see a decrease of parameter $\Delta$ with the latitude increases; the smallest values are observed at solar poles. The symmetry with respect to the ecliptic plane is observed mainly for the fast solar wind, for all magnetic field components measured during solar minima (1997-1998, 2007-2008). This result confirms previous observations~\citep{Bavet00,WawMac10, Wawet15} suggesting similar turbulent properties of the fast polar solar wind in the two hemispheres. On the other hand, we do not have a similar latitudinal dependence and symmetry during the solar maximum (1999-2001), for slow and fast solar wind. The more scattered behavior is observed, as the confirmation of the complex nature of the solar wind during the phase of the strong activity of the Sun.

Let us now  discuss the results obtained for the solar minimum (1997-1998), at heliocentric distance ($\sim 5$ au) and heliographic latitudes smaller than $20^{\circ }$. For these data, we observe that the level of multifractality (intermittency) is lower for fast solar wind than for the slow wind what is in agreement with previous studies performed in the ecliptic~\citep{MarLiu93,Bruet03}, and analysis of \textit{Ulysses} data for the years 1992-1997~\citep{Yoret09}. However, data from solar maximum (1999-2001) and minimum (2007-2008) reveals cases when slow solar wind shows a lower level of multifractality (intermittency) than the fast solar wind. It is worth to recalling that a similar conclusion has been given by~\cite{PagBal02}, from data from the same solar maximum. This is a rather unexpected observation because the fast solar wind is more Alfv\'{e}nic than slow solar wind. The main contribution to solar wind intermittency is due to non-Alfv\'{e}nic structures, while Alfv\'{e}nic increments are found to be characterized by a smaller level of intermittency than the previous ones~\citep[e.g.,][]{Damet12}. To better understand these results we note that most of the slow solar wind cases considered in our analysis pertain to solar cycle 23, which had quite peculiar properties~\citep[e.g.][]{McCet08,Damet11}. In particular, the maximum of solar cycle 23 (the year 2001) was largely dominated by slow wind as expected, which, however, showed a high degree of Alfv\'{e}nicity comparable or even higher than that found in the fast wind during the minimum of the same cycle (2007) as discussed by~\cite{Damet11}. Moreover, recent studies suggest the existence of a new type of  Alv\'{e}nic slow wind having some characteristics common to the fast wind~\citep{DamBru15}. Perhaps, our analysis can be seen as an independent confirmation of these suggestions.

Furthermore, analysis of results presented in Figures~\ref{f:web:latRTN} and~\ref{f:web:latMFRS} reveals some differences between magnetic field components.
In general, in Figure~\ref{f:web:latRTN}(a), (c) and (e), which correspond to fast solar wind data, we observe higher levels of intermittency for the transversal $B_T$ and normal $B_N$ components, for which most of the values of parameter $\Delta$ are in the range of about $0.4$-$0.8$, while in the case of $B_R$ most results present level of multifractality $\Delta$ below $0.6$. This observation confirms previous analyses beyond the ecliptic~\citep{PagBal01,PagBal03,Nicet08}. Differences between multifractality (intermittency) of magnetic components in the case of the slow solar wind (Figure~\ref{f:web:latRTN} (b), (d) and (f)) are rather negligible, with the values of parameter $\Delta$ smaller than $0.6$, in general. Results for data transformed to the MF reference system, as shown in Figure~\ref{f:web:latMFRS}, also reveal differences between components $B_{\parallel}$ (panel (a)), $B_{\perp 1}$ (panels (c)) and $B_{\perp 2}$ (panel (e)). The slightly higher level of intermittency is observed for perpendicular components, which also present some differences between each other, especially for positive latitudes. The lack of the expected significant differences between these components, in particular, the more intermittent state of the $B_{\parallel}$, as shown by~\cite{Bruet03}, can be connected with rather weak contributions of compressive effects on the turbulent behavior. This effect seems to be stronger on smaller scales as we have shown in the previous subsection. 

\subsection{Maps of Multifractality (Intermittency)}
\label{sub:web:map}
 \begin{figure}[!h]
 \centering
\includegraphics[scale=0.95]{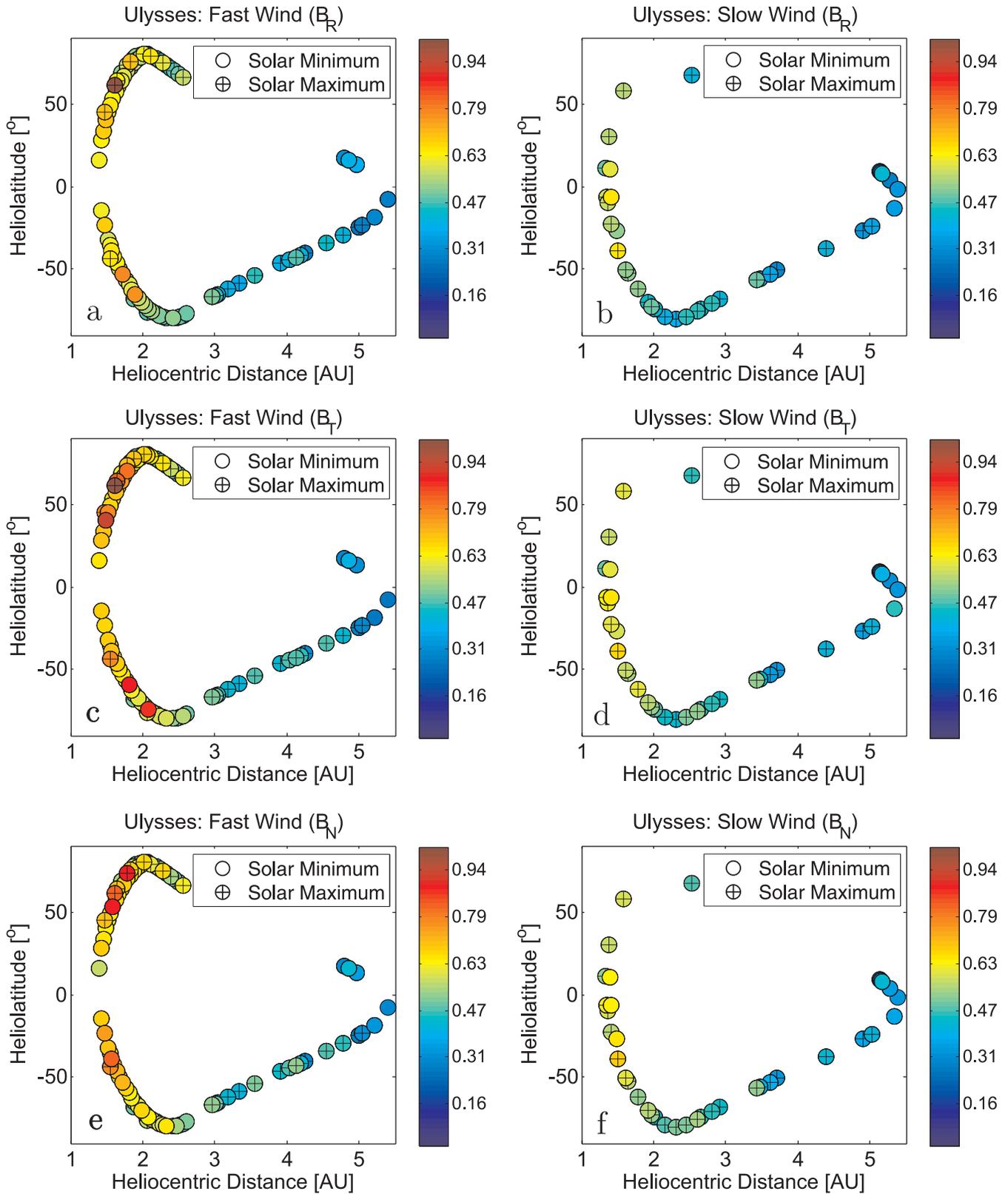}
 \caption{Maps of the degree of multifractality as a measure of intermittency determined for fast (left panel) and slow (right panel) solar wind  system during solar minima (1997-1998, 2007-2008) and solar maximum (1999-2001), respectively. Color denotes the values of the parameter $\Delta$ determined for three magnetic field components: $B_{R}$ (a), (b), $B_{T}$ (c), (d), $B_{N}$ (e), (f) in RTN reference system at different heliocentric distances and heliographic latitudes.}
 \label{f:web:map_RTN}
 \end{figure}

 \begin{figure}[!h]
 \centering
\includegraphics[scale=0.95]{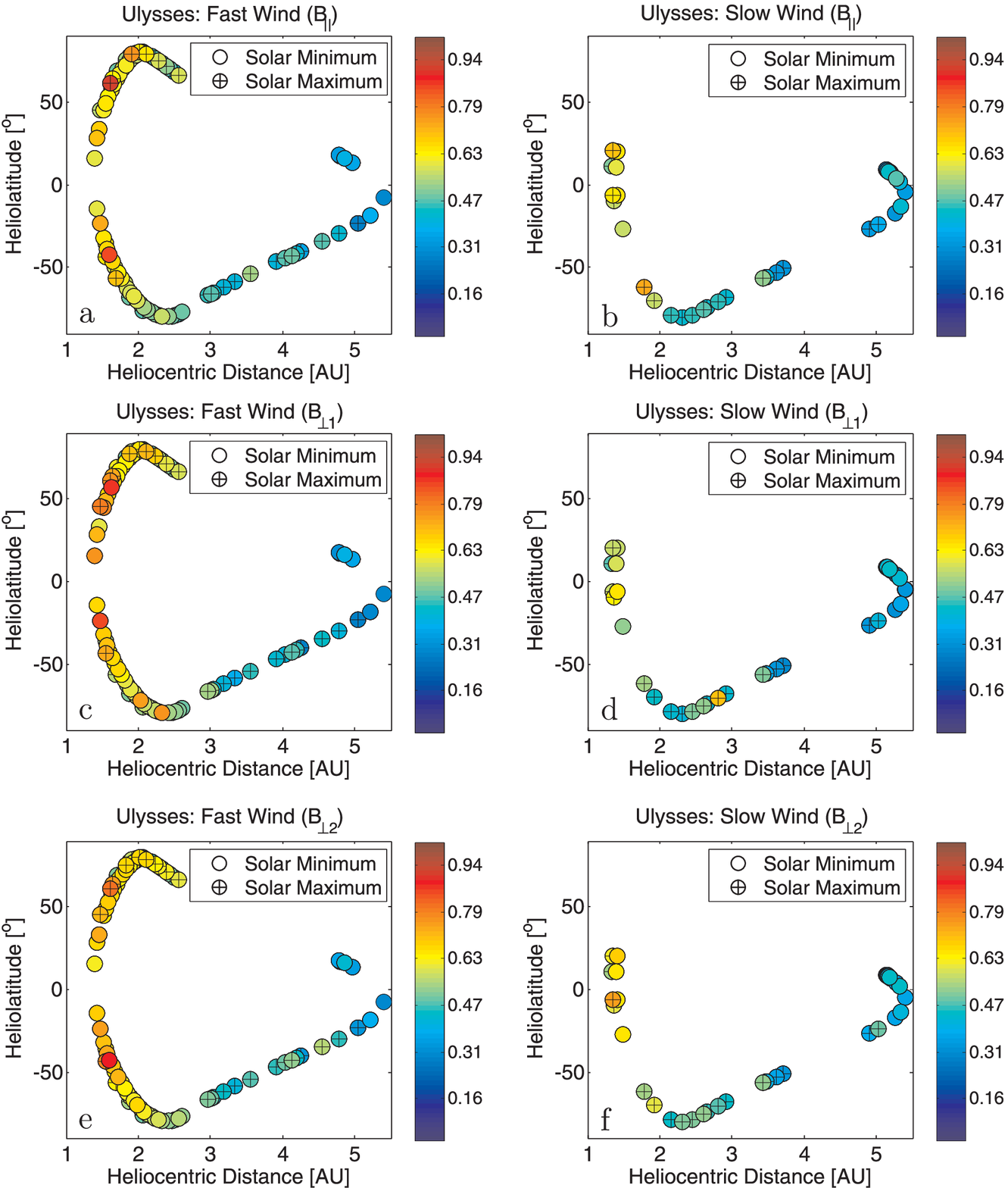}
 \caption{Maps of the degree of multifractality as a measure of intermittency determined for the fast (left panel) and slow (right panel) solar wind  system during solar minima (1997-1998, 2007-2008) and solar maximum (1999-2001), respectively. Color denotes the values of the parameter $\Delta$ determined for the components $B_{\parallel}$ (parallel to the mean field vector $\textbf{B}_0$, panels (a) and (b)), and two components perpendicular to the field: (1) $B_{\perp 1}$ (panels (c) and (d)), $B_{\perp 2}$ (panels (e) and (f)), as observed by \textit{Ulysses} at different heliocentric distances and heliographic latitudes.} 
 \label{f:web:map_MFRS}
 \end{figure}
 
In view of the results of the previous subsections, especially for the second minimum (2007-2008), marked in Figures~\ref{f:web:radRTN} - \ref{f:web:latMFRS} by black circles, we see that the effect of change of solar wind multifractality with heliographic latitude may partially interfere with the variation of this parameter with distance from the Sun. However, it is difficult to separate the latitudinal and radial evolution of intermittency in Ulysses data. Therefore we choose to represent the level of multifractality as a function of both heliocentric distance and heliographic latitude. Figures~\ref{f:web:map_RTN} and~\ref{f:web:map_MFRS} show maps of multifractality along Ulysses orbit, where color denotes the values of the parameter $\Delta$ determined for magnetic field fluctuation in RTN and MF reference systems, respectively. 
The maps clearly show that the fluctuations of the magnetic field in the fast solar wind are most intermittent at small distances from the Sun and this observation is generally independent of the considered magnetic field component. In the case of the slow solar wind, this effect is less significant, differences between components are less obvious. Our analysis shows that the intermittency depends more strongly on the structure of the solar wind than the phase of the solar cycle~\citep{PagBal02}. Moreover, the analysis of \textit{Ulysses'} magnetic field data suggests a somewhat smaller degree of multifractality, $\Delta<1.0$ than obtained previously for velocity fluctuations, $\Delta=1.34-1.52$ \citep{WawMac10}.

\section{Discussions and Conclusions}
\label{sec:web:con}

In this paper, we have studied the scaling properties of high-resolution \textit{Ulysses} magnetic field
fluctuations, measured beyond the ecliptic plane at a wide range of heliocentric distances between $1.4$ and $5.4$ au, at heliolatitudes from $80^{\circ}$S to $80^{\circ}$N. In order to avoid mixing of different physical conditions, we performed a systematic selection of a large number of cases of fast and slow solar wind during two solar minima (1997-1998, 2007-2008) and one solar maximum (1999-2001). The analysis provided about $750$ multifractal spectra that helped to describe the intermittency in the MHD regime (at scales larger than $16$ s) for all magnetic field components in two reference systems: RTN and MF. The main findings of this work can be summarized as follows.

Previous analysis of intermittency in the ecliptic showed the increase with radial distance. Our study points toward a slow decrease of degree of multifractality as a measure of intermittency with distance and with latitude, as initially suggested in a previous analysis of magnetic field intensity~\citep{Wawet15}. This behavior is observed in all magnetic field components, regardless of the reference system used (RTN or MF). The higher rate of intermittency decrease has been found for the fast solar wind, for transversal components in RTN and perpendicular components in the MF reference system. The smallest scales significantly influence on the values of the variation rates.
In general our analysis seems to support the idea that the intermittency in MHD range has a solar origin~\citep{PagBal02, Wawet15}, while the reduced efficiency of drivers (fast and slow streams, shocks interaction, pressure balanced, almost incompressible current sheets, and interplanetary shocks as suggested by~\cite{VelMan99}) does not allow to maintain the level of intermittency with the distance from the Sun.  Additionally, analysis of intermittency over a large range of heliographic latitudes confirmed similar intermittent properties of the fast solar wind turbulence observed in the two hemispheres, as reported in previous studies~\citep{Bavet00,WawMac10,Wawet15}.

The comparison of the slow solar wind intermittency measured during the solar minimum (1997-1998) at distances $\sim 5$ au and close to the equatorial plane presents a higher level of intermittency than the fast solar wind. This confirms previous results~\citep[e.g.,][]{Bruet03,Yoret09} and the need for the separate analysis of the two types of wind. However, analysis of the slow solar wind from solar maximum (1999-2001) and from the second solar minimum (2007-2008) showed in many cases a smaller level of the intermittency than for the fast solar wind. The observation of a high degree of Alfv\'{e}nicity for slow solar wind during solar maximum (2001), as found by~\cite{Damet11}, seems to explain this unusual behavior, as well as the much higher level of multifractality (intermittency) in data from the solar minimum (2007-2008). However, only a detailed analysis of Alfv\'{e}nicity, as well as the further improvement of the data selection procedure by considering additional parameters~\citep[see,e.g.,][]{Lanet12, XuBor14}, especially in the context of classification of the slow solar wind in more than two categories~\citep{DamBru15}, will allow a more detailed verification of this hypothesis. 

Our study suggests that transverse components exhibit a slightly higher level of intermittency than the radial component, similarly to other studies devoted to turbulence beyond the ecliptic~\citep[e.g.,][]{PagBal01, PagBal03}. We noticed relatively small differences between intermittency of parallel and perpendicular components. This difference increased when scales less than $16$ s were included in the analysis, possibly due to increased intermittency at kinetic scales.~\cite{Bruet99} proved that the observed level of magnetic anisotropy strongly depends on intermittency.  Our analysis seems to confirm this fact. In particular, the highest level of anisotropy is observed close to the Sun, where intermittency is the strongest.  However, because we do not find an increase of intermittency with the radial distance, we do not observe an increase of differences between magnetic field components, contrary to what was found in the ecliptic from \textit{Helios} data~\citep[e.g.,][]{Bruet99,Bruet03}.

Finally,~\cite{Soret17} pointed out that different measures of intermittency capture different aspects of fluctuations topology. The multifractal description of the solar wind intermittency based on PF function provides an important complement to previous analyses. In particular, multifractal analysis considers higher-order moments magnifying large or small concentrations of measure, while in the case of basic methods devoted to intermittency only positive or particular moments are used. Therefore, the multifractal spectrum and the degree of multifractality determined in this paper at small scales, provide a complete quantitative and qualitative description of the structure of intermittent magnetic field fluctuations. 
Nevertheless, we are convinced that the statistical description, in particular, the fourth moment of PDF, can still provide an appropriate complementary tool to help us understand the still puzzling aspects of intermittent turbulence beyond the ecliptic.

In summary, the comparison of the intermittent signatures of the magnetic field fluctuations at two very different levels of solar activity over a large range of heliographic latitudes confirms the nonuniversal and the complex nature of solar wind turbulence, and underline the importance of selecting homogeneous samples when performing the analysis.
 
\acknowledgments
We would like to thank  W. M. Macek for the discussion with data selection process.
We would like to thank the SWOOPS, SWICS, and MAG instruments teams of \textit{ULYSSES} mission for providing data. 
This work was supported by the European Community's Seventh Framework Programme ([FP7/2007-2013]) under grant agreement No. 313038/STORM.
M.M.E. acknowledges support from the Belgian Solar Terrestrial Center of Excellence (STCE), the Romanian Space Agency (through project STAR 182-OANA), and the Romanian Ministry of Research (through PCCDI project VESS).

\bibliographystyle{aasjournal}
\bibliography{references-bib-2018}
\end{document}